\documentclass[fleqn,10pt]{wlscirep}
\usepackage[utf8]{inputenc}
\usepackage[T1]{fontenc}

\usepackage{graphicx} 

\usepackage{amsmath,amsthm,amssymb,mathrsfs}
\usepackage{bm}

\usepackage{color}

\title{Input-driven chaotic dynamics in vortex spin-torque oscillator}

\author[1]{Yusuke Imai}
\author[2]{Kohei Nakajima}
\author[1,3]{Sumito Tsunegi}
\author[1,*]{Tomohiro Taniguchi}
\affil[1]{National Institute of Advanced Industrial Science and Technology (AIST), Research Center for Emerging Computing Technologies, Tsukuba, Ibaraki 305-8568, Japan}
\affil[2]{Graduate School of Information Science and Technology, The University of Tokyo, Bunkyo-ku, 113-8656 Tokyo, Japan}
\affil[3]{PRESTO, Japan Science and Technology Agency (JST), Saitama 332-0012, Japan}

\affil[*]{tomohiro-taniguchi@aist.go.jp}

\begin{abstract}
A new research topic in spintronics relating to the operation principles of brain-inspired computing is input-driven magnetization dynamics in nanomagnet. 
In this paper, the magnetization dynamics in a vortex spin-torque oscillator (STO) driven by a series of random magnetic field are studied through a numerical simulation of the Thiele equation. 
It is found that input-driven synchronization occurs in the weak perturbation limit, as found recently. 
As well, chaotic behavior is newly found to occur in the vortex core dynamics for a wide range of parameters, where synchronized behavior is disrupted by an intermittency. 
Ordered and chaotic dynamical phases are examined by evaluating the Lyapunov exponent. 
The relation between the dynamical phase and the computational capability of physical reservoir computing is also studied. 
\end{abstract}

\begin{document}

\flushbottom
\maketitle
%
%


The ability to excite chaotic dynamics in nanomagnets \cite{watelot12,devolder19,bondarenko19,montoya19,williame19,yamaguchi19,taniguchi19PRB,taniguchi19JMMM,kamimaki21} would potentially have practical applications, such as encoding and random number generators \cite{strogatz01}. 
So far, however, it has proven difficult to achieve; 
chaos is a non-periodic and sustainable motion of a physical body maintained by energy injected into a nonlinear system, whereas the magnetization dynamics induced by applying magnetic field and/or current are mainly in the form of switching or a periodic oscillation \cite{katine00,kiselev03,kubota05,tulapurkar05,kubota08,sankey08}. 


Here, several techniques have recently been proposed and experimentally demonstrated with the goal of exciting and identifying chaos, including ones using a delayed-feedback circuit \cite{williame19,taniguchi19PRB,kamimaki21} and ones that add another ferromagnet \cite{taniguchi19JMMM,devolder19,bondarenko19,montoya19,williame19,yamaguchi19}. 
Another possible stage on which to demonstrate chaotic dynamics is an input-driven dynamical system \cite{manjunath12}, where a series of time-dependent stimuli drive the nonlinear dynamics of a physical system. 
An example that utilizes such input-driven dynamics is human-voice recognition by a vortex spin-torque oscillator (STO) \cite{torrejon17}, which is based on the algorithm of physical reservoir computing \cite{maas02,jaeger04,verstraeten07}. 
Wherein a human voice is converted into an electric-voltage input signal that modulates the dynamical amplitude of the vortex core. 
The input signal can be recognized when a one-to-one correspondence between it and the vortex dynamics is obtained through machine learning. 
Such a correspondence can be regarded as synchronization between the input signal and the STO response. 
A critical difference between this sort of synchronization and the conventional synchronizations studied in STOs previously, such as forced synchronization \cite{quinsat11,rippard13}, is that the input signal is generally non-periodic, whereas the input in the conventional cases usually has a certain periodicity. 
Moreover, while the STO dynamics in the conventional cases become identical to the input signal except a phase difference, the STO dynamics in this case are not generally identical to the input signal but have a certain reproducibility; the STO response is the same when the same input signal is injected. 
Thus, it is a wider concept of synchronization that encompasses the conventional ideas, which can be called generalized synchronization \cite{pikovsky03}, i.e., is a kind of input-driven dynamics first examined in the 1990s \cite{pecora90,maritan94,rulkov95}. 
Another example of input-driven dynamics is chaos \cite{crutchfield82,akashi20}, near which the computational capability of physical reservoir computing is sometimes enhanced \cite{bertschinger04,nakayama16}.  
As can be seen from these examples, input-driven dynamics is of great interest to researchers studying brain-inspired computing, because its complexity can be used for information processing \cite{nakajima20,nakajima21}. 
However, there are only a few reports \cite{akashi20,yang07,akashi22,taniguchi22} on input-driven dynamics, in particular chaos, in spintronics devices.

In this work, we study input-driven magnetization dynamics in a vortex STO by performing a numerical simulation of the Thiele equation. 
A uniformly distributed random magnetic field is used as a series of input signals; such input signals have often been used to study of input-driven dynamics \cite{kubota21}. 
We found that when the input is weak, input-driven synchronization occurs\cite{imai22}; the output of the STO is determined by the input signal and is independent of the initial state. 
On the other hand, in the strong input limit, input-driven chaotic dynamics appear, in the form of an intermittency in the synchronized behavior, and therefore, the temporal dynamics depend on the initial state. 
We distinguish these ordered and chaotic dynamical phases by evaluating the largest conditional Lyapunov exponent, which is a measure quantifying the sensitivity of the dynamics to the initial state of the STO.  
We also study the relation between the dynamical phase and computational capability of the STO by relating the Lyapunov exponent to the short-term memory capacity.


\section*{Model}

\subsection*{Thiele equation}
\begin{figure}[h]
\centerline{\includegraphics[width=1.0\columnwidth]{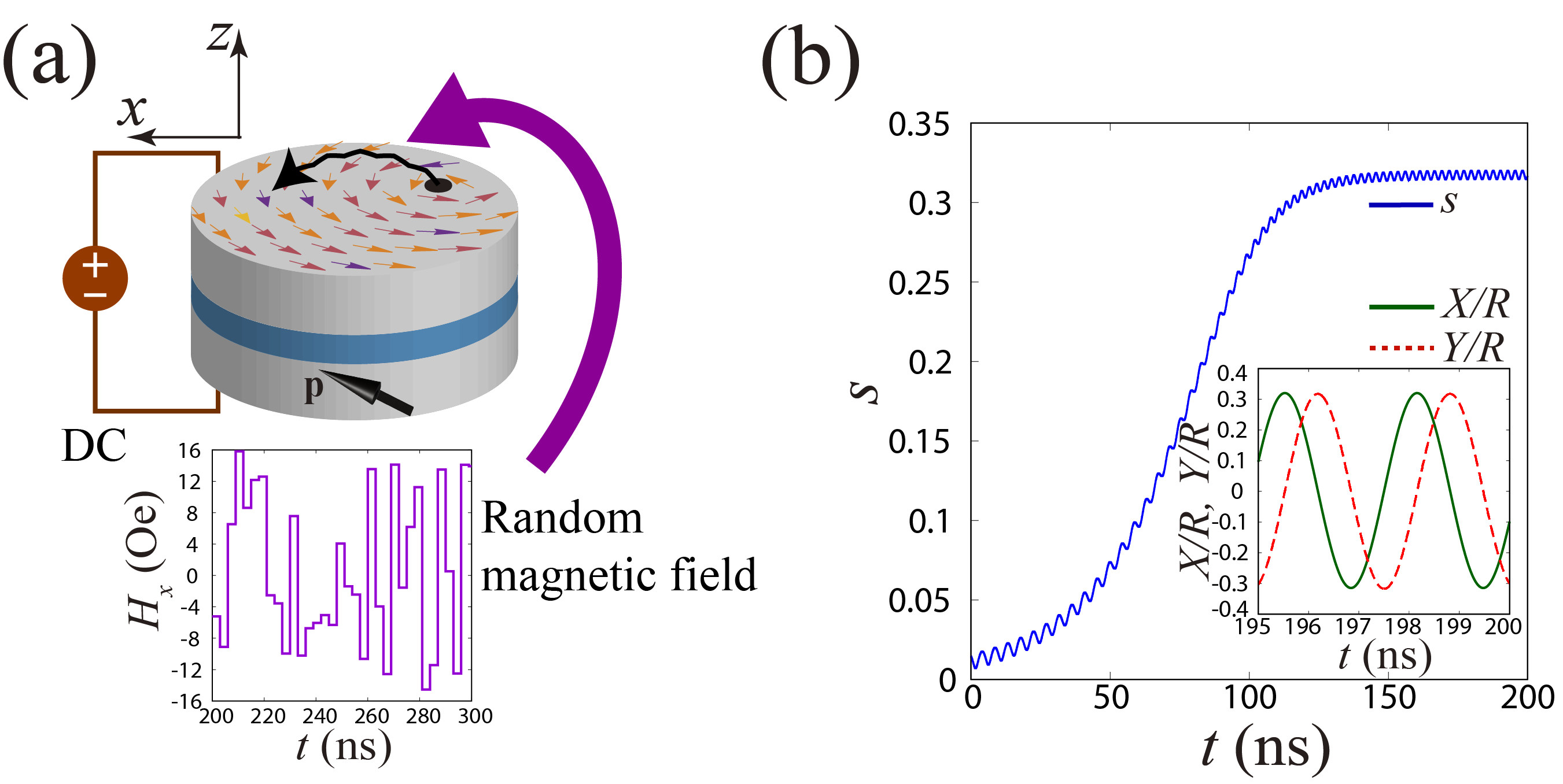}}
\caption{
             (a) Schematic illustration of vortex spin torque oscillator and dynamics of the vortex core induced by a direct current (DC) and random magnetic field. 
                 The DC induces an oscillation of the vortex core around the disk center whereas the random magnetic field modulates the oscillation amplitude. 
             (b) Evolution of the normalized amplitude $s$ in the presence of the current $I=5$ mA. 
                 The inset shows the dynamics of the $x$ component and $y$ component of the magnetic vortex core normalized by the radius of the magnetic vortex core.
         \vspace{-3ex}}
\label{fig:fig1}
\end{figure}

The system we consider is schematically shown in Fig. \ref{fig:fig1}(a), where the vortex STO consists of a ferromagnetic/nonmagnetic/ferromagnetic trilayer. 
The free layer has a magnetic vortex, while the reference layer is uniformly magnetized. 
The $z$ axis is normal to the film-plane, while the $x$ axis is parallel to the projection of the magnetization in the reference layer onto the film-plane. 
Throughout this paper, the unit vector pointing in the direction of the $k$-axis ($k=x,y,z$) is denoted as $\mathbf{e}_{k}$. 
It has been shown that the experimental and numerical results on the magnetization dynamics in a vortex STO are well described by the Thiele equation,  \cite{thiele73,guslienko06PRL,guslienko06,ivanov07,khvalkovskiy09,guslienko11,dussaux12,grimaldi14}
\begin{equation}
\begin{split}
  &
  -G \mathbf{e}_{z}
  \times
  \dot{\mathbf{X}}
  -
  |D|
  \left(
    1
    +
    \xi 
    s^{2}
  \right)
  \dot{\mathbf{X}}
  -
  \kappa
  \left(
    1
    +
    \zeta 
    s^{2}
  \right)
  \mathbf{X}
  +
  a_{J} J p_{z}
  \mathbf{e}_{z}
  \times
  \mathbf{X}
  +
  c a_{J} J R_{0} 
  p_{x}
  \mathbf{e}_{x}
  +
  c \mu^{*}
  \mathbf{e}_{z}
  \times
  \mathbf{H}
  =
  \bm{0}, 
  \label{eq:Thiele}
\end{split}
\end{equation}
where $\mathbf{X}=(X,Y,0)$ is the position vector of the vortex core, while $G=2\pi pML/\gamma$ and $D=-(2\pi\alpha ML/\gamma)[1-(1/2)\ln(R_{0}/R)]$ consist of the saturation magnetization $M$, the gyromagnetic ratio $\gamma$, the Gilbert damping constant $\alpha$, the thickness $L$, the disk radius $R$, and the core radius $R_{0}$ of the free layer. 
The polarity $p$ and chirality $c$ are each assumed to be $+1$ for convenience. 
The normalized distance of the vortex core's position from the disk center is $s=|\mathbf{X}|/R=\sqrt{(X/R)^{2}+(Y/R)^{2}}$. 
The dimensionless parameter $\xi$ quantifies the nonlinear damping in a highly excited state \cite{dussaux12}.
The parameters $\kappa$ and $\zeta$ relate to the magnetic potential energy $W$ via 
\begin{equation}
  W
  =
  \frac{\kappa}{2}
  \left(
    1
    +
    \frac{\zeta}{2}
    s^{2}
  \right)
  |\mathbf{X}|^{2}, 
  \label{eq:potential_energy}
\end{equation}
and $\kappa=(10/9)4\pi M^{2}L^{2}/R$ \cite{dussaux12}. 
The dimensionless parameter $\zeta$ provides the linear dependence of the oscillation frequency on the current \cite{dussaux12}. 
As mentioned below, the oscillation amplitude of the vortex core is determined by the current magnitude, and thus, we call $\zeta$ the amplitude-frequency coupling parameter, for convenience. 
The spin-transfer torque strength with spin polarization $P$ is $a_{J}=\pi\hslash P/(2e)$ \cite{khvalkovskiy09,guslienko11}.
The electric current density $J$ is positive when the current flows from the reference to the free layer. 
The current is denoted as $I=\pi R^{2}J$. 
The unit vector pointing in the magnetization direction in the reference layer is $\mathbf{p}=(p_{x},0,p_{z})$. 
The external magnetic field is denoted as $\mathbf{H}$, while $\mu^{*}=\pi MLR$. 
The material parameters are summarized in Methods. 
Note that the current is an experimentally controllable variable; typically, it is on the order of $1$ mA ($I < 10$ mA \cite{tsunegi21}). 
Moreover, the value of $\zeta$ relates to the material and is on the order of $0.1$-$1.0$ \cite{grimaldi14}. 
Therefore, we will study the dynamical behavior of the vortex core as a function of $J$ (or equivalently, $I$) and $\zeta$; see also Methods. 

In the absence of a current and magnetic field, the vortex core is located near the disk center. 
By applying a direct current, spin-transfer torque moves the vortex core from the disk center and excites an auto-oscillation with an approximately constant amplitude when the current density is larger than a critical value \cite{yamaguchi20}, 
\begin{equation}
  J_{\rm c}
  =
  \frac{|D|\kappa}{Ga_{J}p_{z}}.
  \label{eq:Jc}
\end{equation}
The oscillation amplitude and frequency are $s_{0}=\sqrt{[(J/J_{\rm c})-1]/(\xi+\zeta)}$ and $f_{0}=(\kappa/G)(1+\zeta s_{0}^{2})/(2\pi)$, respectively. 
The condition that the vortex core is inside the ferromagnetic disk, i.e., $|\mathbf{X}|<R$ or equivalently, $s < 1$, implies that the Thiele equation is applicable when the current density is less than another threshold value, 
\begin{equation}
  J_{\rm th}
  =
  J_{\rm c}
  \left(
    1
    +
    \xi
    +
    \zeta
  \right).
  \label{eq:Jth}
\end{equation}
Figure \ref{fig:fig1}(b) shows an example of such an auto-oscillation, where $I=5.0$ mA and $\zeta=7.52$. 
The figure indicates that the vortex core moves from the disk center ($s=0$) and shows an oscillation around the center with an approximately constant radius $s_{0}$. 
The small-amplitude oscillation of $s$ is due to the fifth term in Eq. (\ref{eq:Thiele}), which originates from the in-plane component of $\mathbf{p}$ and breaks the axial symmetry of the Thiele equation around the $z$ axis.

The input-driven dynamics appear when a series of time-dependent signals is injected to this auto-oscillation state. 
The input signal in this study is a random-pulse magnetic field, as in an experiment on physical reservoir computing \cite{tsunegi19}, 
\begin{equation}
  \mathbf{H}(t)
  =
  h_{x} r(t) 
  \mathbf{e}_{x}, 
  \label{eq:input}
\end{equation}
where $h_{x}$ is the maximum value of the input magnetic field, and $r(t)$ is a uniformly distributed random number in the range of $-1 \le r(t) \le 1$ whose corresponding input signal has a pulse width $t_{\rm p}$. 
Although the pulse input signal in the experiments has a rise time, it can be shortened by the preprocessing \cite{tsunegi18JJAP}. 
For simplicity, we assume a step-function pulse in this work.


\section*{Results}


\subsection*{Temporal dynamics}

\begin{figure}[h]
\centerline{\includegraphics[width=1.0\columnwidth]{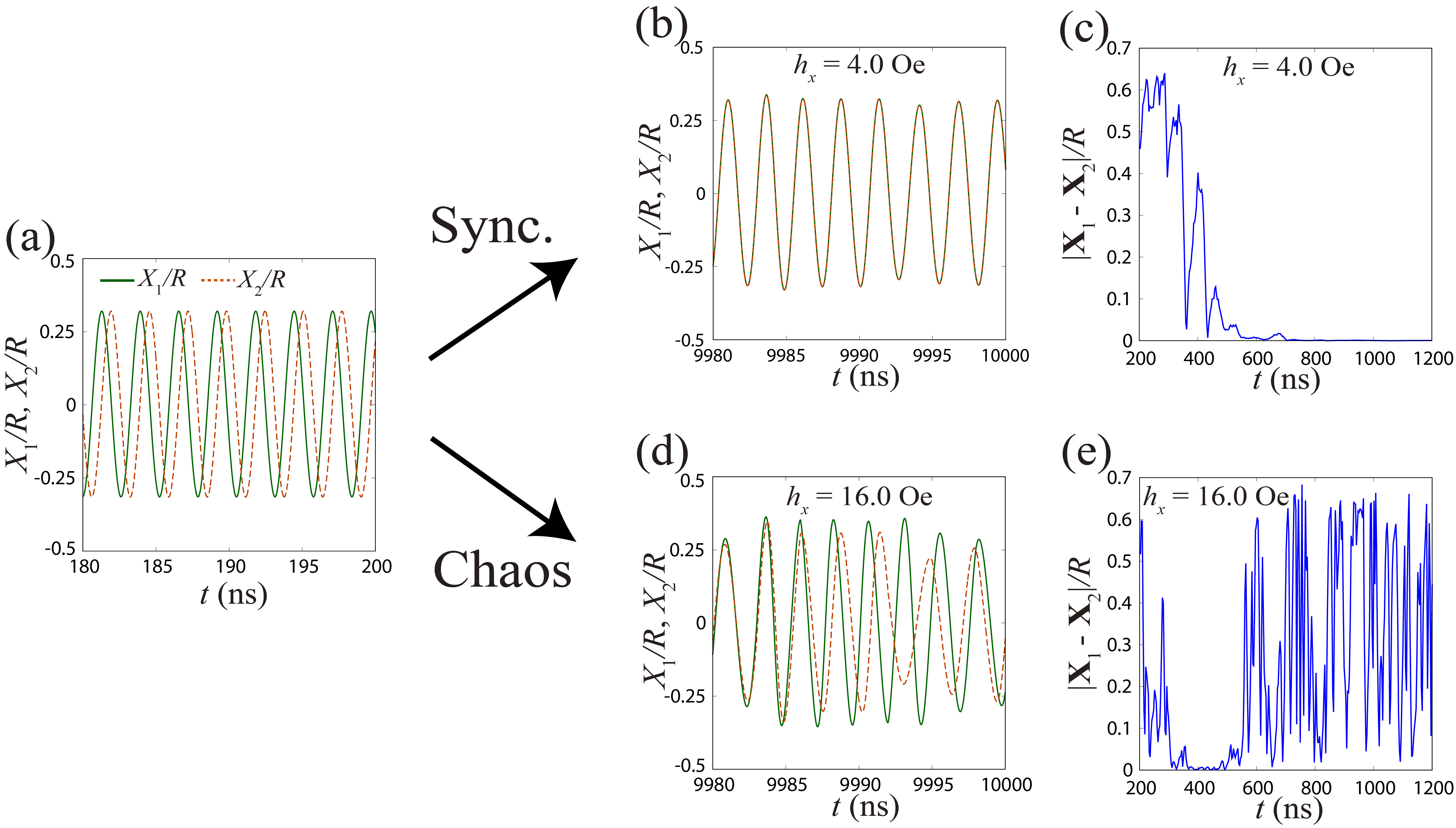}}
\caption{
            (a) Temporal dynamics of two $X/R$s, denoted as $X_{1}/R$ and $X_{2}/R$, with slightly different initial condition. 
                 Here, the random magnetic field is not injected yet. 
            (b) Temporal dynamics of $X_{1}/R$ and $X_{2}/R$ after injection of the input signal.  
                 The input signal is relatively weak ($h_{x}=4$ Oe). 
            (c) Difference between $\mathbf{X}/R$s, $|\mathbf{X}_{1}-\mathbf{X}_{2}|/R$, for a small input strength. 
            (d) Temporal dynamics of $X_{1}/R$ and $X_{2}/R$ after injection of the input signal.  
                 The input signal is relatively strong ($h_{x}=16$ Oe). 
            (e) Difference between $\mathbf{X}/R$s, $|\mathbf{X}_{1}-\mathbf{X}_{2}|/R$, for a large input strength. 
         \vspace{-3ex}}
\label{fig:fig2}
\end{figure}

The input magnetic field given by Eq. (\ref{eq:input}) causes two kinds of dynamics; synchronization and chaotic. 
Figure \ref{fig:fig2}(a) shows two solutions of Eq. (\ref{eq:Thiele}) with slightly different initial conditions before injection of the input signal, where $I=5.0$ mA and $\zeta=7.52$. 
For convention, let us denote the two solutions as $\mathbf{X}_{1}$ and $\mathbf{X}_{2}$. 
The solutions oscillate with the same frequency but different phases. 
When a weak input signal is injected from $t=200$ ns, they overlap in the limit $t\to \infty$; i.e., synchronization occurs; see Fig. \ref{fig:fig2}(b), where $h_{x}=4.0$ Oe and $t_{\rm p}=3.0$ ns. 
The difference between the two solutions, $|\mathbf{X}_{1}-\mathbf{X}_{2}|/R$, is illustrated in Fig. \ref{fig:fig2}(c). 
As shown, the difference decreases over time, which also indicates that phase synchronization is achieved. 
Recall that the input signal does not have periodicity; therefore, the synchronization of the two solutions is different from the forced synchronization found in, for example, Refs. \cite{quinsat11,rippard13}. 
The synchronization here is input-driven \cite{mainen95,toral01,teramae04,goldobin05,nakao07}, where the temporal dynamics are determined by the input signal, as was found in a vortex STO recently \cite{imai22}. 
It occurs when the strength of the input signal is relatively weak. 

Chaotic dynamics also appear as the input-signal strength increases. 
Figure \ref{fig:fig2}(d) shows an example with $h_{x}=16.0$ Oe, where the two solutions of Eq. (\ref{eq:Thiele}) were arrived at with slightly different initial conditions.  
As shown, the solutions are never completely overlapped even though a long time passes. 
The difference between the two solutions often becomes small; i.e., synchronized behavior appears. 
However, before the difference completely becomes zero, intermittency appears and the two solutions separate from each other; see also Fig. \ref{fig:fig2}(e).  
This result indicates that the dynamical behavior of the vortex core is not solely determined by the input signal, but is also sensitive to its initial state. 
The sensitivity of the dynamical behavior implies chaos. 
We emphasize that the chaotic behavior is caused by the input signal. 
In fact, without the input signal, the system described by the Thiele equation is a two-dimensional autonomous system, and in such a system, chaos is prohibited by Poincar\'e-Bendixson theorem \cite{alligood97}. 
Therefore, the behavior of the vortex core must be input-driven chaos. 

The chaotic dynamics found here derives from the temporal dynamics. 
To identify it more systematically, we can evaluate the Lyapunov exponent.


\subsection*{Lyapunov exponent}

The Lyapunov exponent is the expansion rate of the difference between two solutions of an equation of motion with slightly different initial conditions. 
In the present case, we will define the difference between two solutions, $\mathbf{X}_{1}$ and $\mathbf{X}_{2}$, as $\delta x = |\mathbf{X}_{1}-\mathbf{X}_{2}|/R$. 
When $\mathbf{X}_{1}\simeq \mathbf{X}_{2}$ at a certain time $t_{0}$, the time development of $\delta x$ can be described by a linear equation, $\delta\dot{x}\simeq \lambda \delta x$. 
The solution is $\delta x(t)\simeq \delta x(t_{0}) e^{\lambda t}$, and $\lambda$ is the Lyapunov exponent. 
Accordingly, the two solutions become identical when the Lyapunov exponent is negative. 
The dynamics corresponding to a negative Lyapunov exponent are, for example, saturation to a fixed point and phase synchronization. 
On the other hand, when the Lyapunov exponent is positive, the difference between the two solutions increases, indicating sensitivity of the dynamics to the initial state. 
A positive exponent implies the appearance of chaos. 
Moreover, when the exponent is zero, the initial difference is kept constant. 
An example of such dynamics is a limit-cycle oscillation. 
The method of evaluating the Lyapunov exponent is summarized in Methods. 

\begin{figure}[h]
\centerline{\includegraphics[width=1.0\columnwidth]{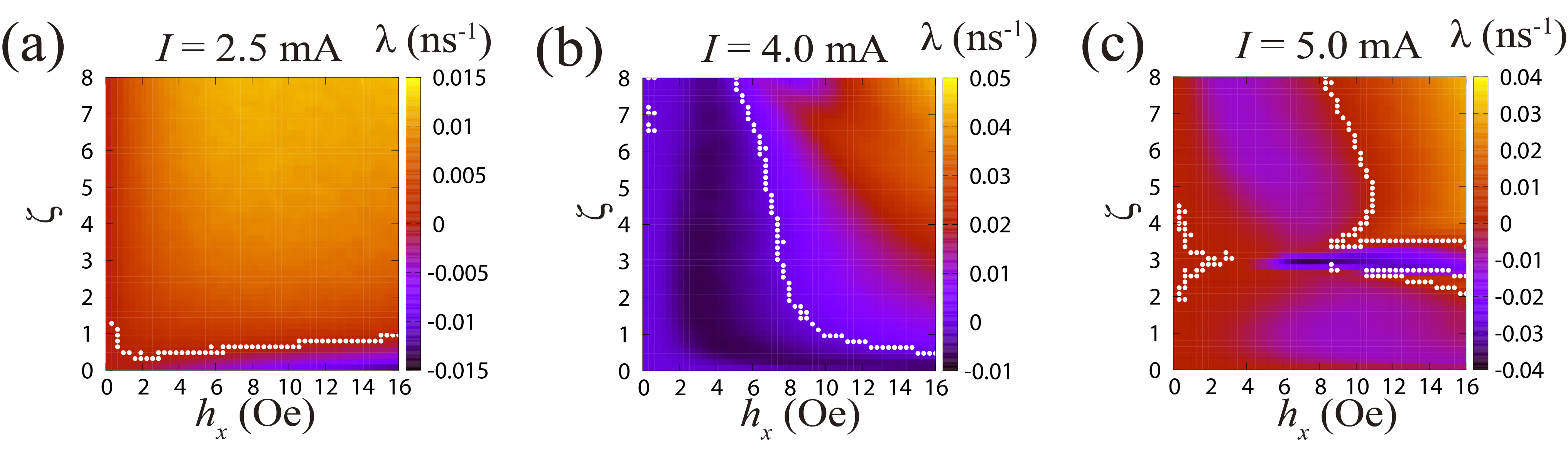}}
\caption{ 
             Dependence of Lyapunov exponent ($\lambda$) on the strength of the input signal ($h_{x}$) and amplitude-frequency coupling parameter ($\zeta$) for currents of (a) $I=2.5$ mA, (b) $I=4.0$ mA, and (c) $I=5.0$ mA.
             The pulse width is $3.0$ ns. 
             The white dots represent the zero-exponent regions. 
         \vspace{-3ex}}
\label{fig:fig3}
\end{figure}

Figure \ref{fig:fig3} summarizes the dependence of the Lyapunov exponent on the input magnetic field strength $h_{x}$ and amplitude-frequency coupling parameter $\zeta$ for various current values $I$; (a) $2.5$, (b) $4.0$, and (c) $5.0$ mA. 
The pulse width of the input signal is $t_{\rm p}=3.0$ ns. 
The figure provides a broad perspective on the relation between the parameter values and the dynamical phases, although the detailed structure depends on the current value (see also Supplementary Information, which has the other data supporting the discussion below, obtained for various currents and pulse widths).


First, let us focus on the parameter regions where the values of $h_{x}$ and $\zeta$ are relatively small. 
The Lyapunov exponent takes on large negative values when $h_{x}$ or $\zeta$ increases. 
This tendency is clear in, for example, Fig. \ref{fig:fig3}(b), from $h_{x}\simeq 0$ to $h_{x}\simeq 4$ Oe. 
On the other hand, it seems somewhat unclear for $I=2.5$ mA in Fig. \ref{fig:fig3}(a); but a closer look to small $h_{x}$ and $\zeta$ regions shows the negative enhancement of the exponent; see also Supplementary Information.  
This result implies that the input-induced synchronization is rapidly realized because the negative Lyapunov exponent corresponds to the inverse of the time scale decreasing the difference in the initial state; i.e., a large negative Lyapunov exponent means rapid synchronization. 
When the input strength $h_{x}$ is large, the input signal substantially modifies the vortex-core position and contributes to a fast alignment of $\mathbf{X}_{1}$ and $\mathbf{X}_{2}$. 
Similarly, when $\zeta$ is large, the oscillation frequency of the vortex core strongly depends on its oscillation amplitude $s$, and therefore, the vortex cores having different amplitudes oscillate with different frequencies and decrease their phase difference. 
Accordingly, synchronization easily occurs as $h_{x}$ and $\zeta$ increase, resulting in a large negative Lyapunov exponent. 

Increasing the values of these parameters any further, however, makes it more difficult to fine tune the vortex-core position, because even a single input signal causes a large change in position.  
In such case, it is hard to align the vortex core positions by injection of an input signal; rather, a small difference between the initial states leads to a big difference in position. 
As a result, chaotic dynamics appears, which is confirmed by the positive Lyapunov exponent. 
We should also notice that the dependence of the Lyapunov exponent on the parameters is sometimes non-monotonic. 
For example, for $I=5.0$ mA in Fig. \ref{fig:fig3}(c), at $h_{x}\gtrsim 5.0$ Oe, the Lyapunov exponent becomes negative or positive depending on the value of $\zeta$ around $\zeta=3.0$. 
Such a parameter dependence has been frequently observed in nonlinear dynamical systems showing chaos \cite{strogatz01}; see also Supplemental Information showing the dependence of the Lyapunov exponent on $\zeta$ and the bifurcation diagram. 

In summary, increasing the values of the system parameters leads to an enhancement in the efficiency of achieving input-driven synchronization, which can be quantified by an increasingly large negative Lyapunov exponent. 
Increasing the parameters further, however, results in a transition of the dynamical phase from synchronized to chaotic, because a large modulation of the vortex-core position expands the difference between the initial states. 
Chaotic dynamics appears when the Lyapunov exponent is positive. 

At the end of this section, let us give a brief comment on the verification of chaos experimentally. 
In the experiments, the initial state of the vortex core is uncontrollable. 
Therefore, it is difficult to evaluate the Lyapunov exponent experimentally by a method similar to that used here, although there are statistical methods estimating the Lyapunov exponent from the experimentally observed data \cite{sano85,rosenstein93,kantz94}. 
In addition, the stochastic torque due to thermal fluctuation \cite{brown63} makes the vortex-core dynamics complex, which also makes it difficult to evaluate the Lyapunov exponent. 
Note that there are other methods identifying chaos from the experimentally observed data. 
For example, the noise limit \cite{barahona96,poon01} can be a figure of merit for the identification of chaos; in fact, it has been used in the recent experiments using the vortex STOs \cite{devolder19,kamimaki21}. 
The noise limit has been, however, mainly evaluated in autonomous systems, while the present system is nonautonomous. 
An extension making it applicable to the nonautonomous system might be necessary, which could be a future work. 


\section*{Discussion}

We have found that chaotic dynamics in a vortex STO can be driven by injection of a random magnetic-field pulse. 
Moreover, Ref. \cite{akashi20} found that a pulse-current driven bifurcation from ordered to chaotic dynamics occurs in a macrospin STO. 
While both results were obtained through numerical simulations, our findings, which are based on the use of a magnetic field, might be more suitable for an experimental verification. 
This is because excitation of chaos often requires a large input signal, and a large current can be used for generating a magnetic field, whereas there is a restriction on the magnitude of the current that can be directly applied to an STO because of its high resistance; see also Methods. 

\begin{figure}[h]
\centerline{\includegraphics[width=1.0\columnwidth]{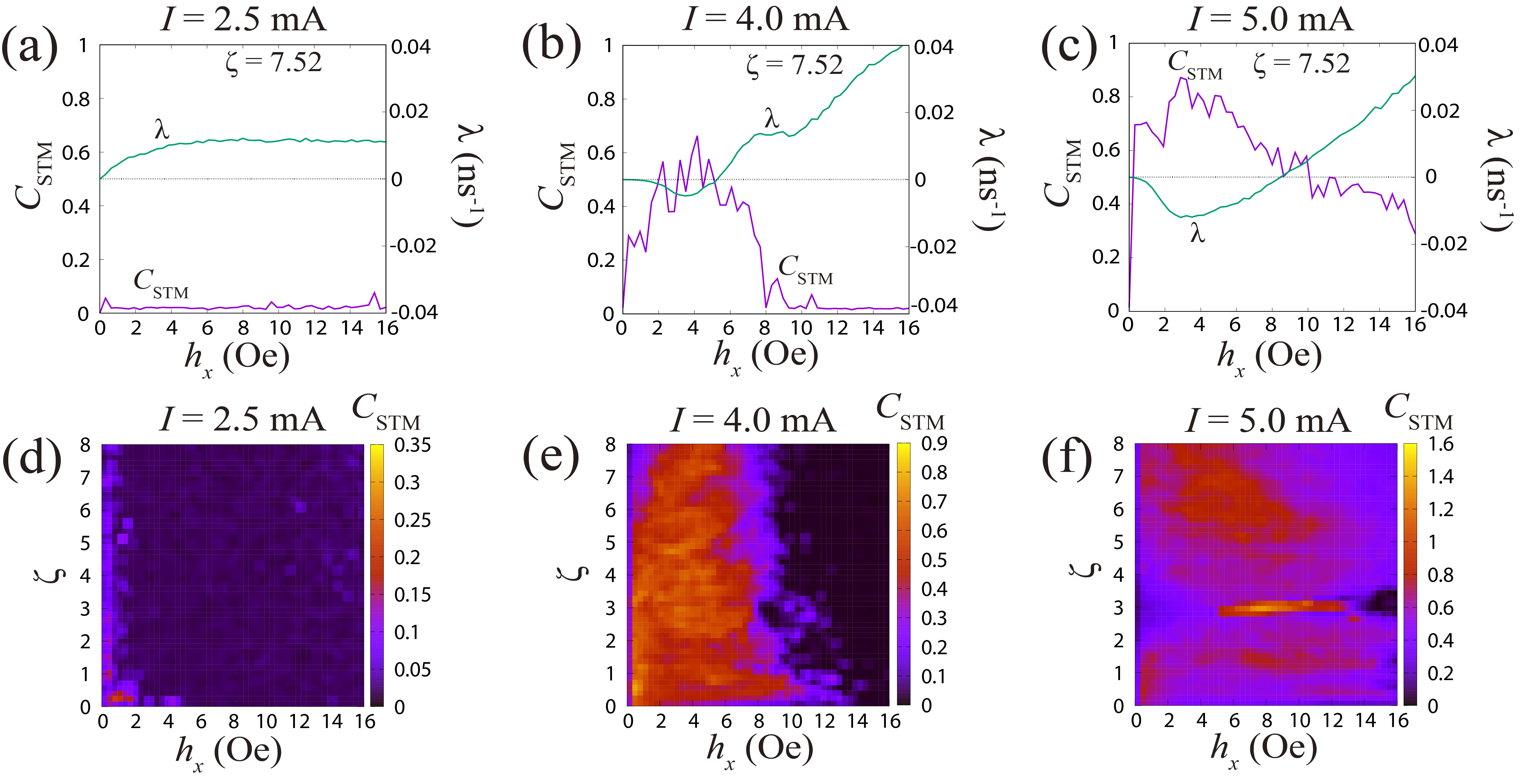}}
\caption{
             Dependence of short-term memory capacity ($C_{\rm STM}$) and the Lyapunov exponent $\lambda$ on the input strength $h_{x}$ for currents $I$ of (a) $2.5$, (b) $4.0$, and (c) $5.0$ mA; here $\zeta=7.52$. 
             Dependence of $C_{\rm STM}$ and $h_{x}$ on $h_{x}$ and $\zeta$ for currents of (d) $2.5$, (e) $4.0$, and (f) $5.0$ mA. 
         \vspace{-3ex}}
\label{fig:fig4}
\end{figure}

As mentioned in the Introduction, an interesting application of input-driven dynamics is information processing. 
Let us, therefore, evaluate the computational capability of the STO and investigate the relation between this capability and the dynamical phase. 
As a figure of merit, we will use the short-term memory capacity \cite{tsunegi18JJAP} of physical reservoir computing, which corresponds to, roughly speaking, the number of random input signals a physical reservoir can recognize; see also Methods describing the evaluation of the memory capacity. 
A large memory capacity corresponds to high computational capability. 
Figure \ref{fig:fig4} plots the value of the short-term memory capacity for various currents $I$ for the pulse width of $3.0$ ns. 
In particular, Fig. \ref{fig:fig4}(a)-\ref{fig:fig4}(c) show the dependence of the short-term memory capacity, as well as that of the Lyapunov exponent, on the input strength $h_{x}$ for a fixed $\zeta$, while Figs. \ref{fig:fig4}(d)-\ref{fig:fig4}(f) show the dependence of the short-term memory capacity on $h_{x}$ and $\zeta$; see also Supplemental Information for the short-term memory capacity obtained for different pulse widths. 
The short-term memory capacity is large when the current magnitude is large. 
It is attributed to the current dependence of the relaxation time. 
The short-term memory capacity becomes large when the reservoir recognizes the input data well. 
The STO recognizes the input data by the change of the oscillation amplitude. 
When the current is small, the relaxation time, or the response of the vortex core, is slow \cite{yamaguchi20}, and therefore, the input data is not well recognized. 
Therefore, the input data is not well recognized by the STO, leading to a small short-term memory capacity. 
On the other hand, when the current is large, the vortex core responds to the input data immediately. 
Therefore, the STO recognizes the injection of the input data leading to a large short-term memory capacity. 
We also note that the magnitude relationship between the relaxation time and the pulse width affects the computational capability, which was studied in Ref. \cite{yamaguchi20srep}. 


Comparing the results shown in Fig. \ref{fig:fig4} with those in Fig. \ref{fig:fig3} leads us to the following conclusions. 
The short-term memory capacity is large in the parameter regions where the Lyapunov exponent is negative. 
On the other hand, it becomes almost zero when the exponent is positive. 
These results can be interpreted as follows. 
A recognition task of an input signal is done by learning the response from the STO. 
Imagine that we inject identical input signals into the STO at different times. 
In general, the initial states of the STO for these different trials are different. 
Therefore, for STO to be able to recognize that input signals are the same, the output signal from it should saturate to a certain state that is independent of the initial ones.  
In other words, the temporal response of the STO should be solely determined by the input signal. 
Note that a negative Lyapunov exponent corresponds to the dynamics saturating to a fixed point and thus becoming independent of the initial state. 
Therefore, the dynamical phase corresponding to a negative Lyapunov exponent is applicable to a recognition task. 
On the other hand, when the dynamics are sensitive to the initial state, the STO will output different signals for the injection of a common input signal. 
In such a case, the success rate of a recognition task will be low. 
Chaos is a typical dynamics having sensitivity to the initial state. 
Therefore, the short-term memory capacity becomes zero when the Lyapunov exponent is positive. 
At the same time, we should note that, although chaotic dynamics are not useful for computing, its dynamical neighborhood can be used for computing. 
In some cases \cite{nakajima21}, the boundary of the dynamical phase is useful for particular tasks. 
For example, the short-term memory capacity is maximized near the edge of chaos in some physical reservoirs \cite{bertschinger04,nakayama16}, although this is not the present case; see also Supplementary Information, where the Lyapunov exponent and the short-term memory capacity for various parameters are summarized. 
Therefore, performing computation even in a chaotic region would help to deepen our understanding of the computational capability of STOs. 


In summary, we showed that input-driven chaotic magnetization dynamics can be excited in a vortex STO in a numerical simulation of the Thiele equation. 
Chaos was identified from the temporal dynamics of the solutions of the equation with slightly different conditions, as well as the analysis of the Lyapunov exponent. 
The correspondence between the dynamical phase of the input-driven STO and the short-term memory capacity of physical reservoir computing was also investigated. 
The results showed that the performance of brain-inspired computing relates to the dynamical phase of the input-driven dynamical systems, and therefore, manipulation of the phase is necessary for practical applications. 
The present work indicates that the dynamical phase of STO can be tuned between input-driven synchronized and chaotic states by system parameters. 
While the present work chose physical reservoir computing as an example, chaotic dynamics might have different applications in brain-inspired computing because chaotic dynamics has been found in artificial neural networks emulating brain activities \cite{aihara90}. 
The investigation of such applications will be a future work.

\section*{Methods}


\subsection*{Material parameters}

The material parameters were taken from typical experiments and simulations \cite{dussaux12,grimaldi14,tsunegi21} as $M=1300$ emu/cm${}^{3}$, $\gamma=1.764\times 10^{7}$ rad/(Oe s), $\alpha=0.01$, $L=5$ nm, $R=187.5$ nm, $R_{0}=10$ nm, $P=0.7$, 
$\xi=2$, and $\mathbf{p}=(\sin 60^{\circ},0,\cos60^{\circ})$. 
A current $I$ of $1$ mA corresponds to a current density $J$ of $0.9$ MA/cm${}^{2}$. 
The value of the current used in, for example, the experiment in Ref. \cite{tsunegi21} is $8$ mA at maximum. 
The value of $\zeta$ in Ref. \cite{grimaldi14} in the absence of a current is defined as $\zeta=\kappa_{\rm ms}^{\prime}/\kappa_{\rm ms}$, where $\kappa_{\rm ms}$ and $\kappa_{\rm ms}^{\prime}$ relate to the magnetostatic energy and their values are determined by fitting experiments and/or numerical analysis of the Thiele equation. 
The values of these parameters in Ref. \cite{grimaldi14} are widely distributed between the experimental (about $\zeta\sim 5$) and theoretical (about $0.2$) analyses. 

A typical magnitude of the input magnetic field used in the present work is on the order of $1$-$10$ Oe. 
In experiments, a pulse magnetic field can be generated by applying a pulse electric current to a metal line. 
For example, assuming a thin metal line, a current of $1$ mA flowing above an STO within a distance on the order of $10$ nm generates a magnetic field on the order of $1$-$10$ Oe, where we have used the Amp\'ere law, $H=I/(2\pi \mathscr{R})$ with current $I$ and distance $\mathscr{R}$. 
In reality, the metal line has a finite width and the magnitude of the magnetic field will be small; 
for example, in Ref. \cite{suto17}, a metal line is placed on a nanomagnet within a distance of $20$ nm, and current of $20$ mA generates a magnetic field of $80$ Oe. 
Note that the magnitude of the magnetic field could be further enhanced by applying a large current, which is possible because the current is in a metal line. 
On the other hand, Ref. \cite{akashi20} reported a bifurcation from order to chaos by injection of electric current pulses into a macrospin STO, where the current magnitude was on the order of $1$-$10$ mA. 
However, it should be noted that a large current, such as $10$ mA, leads to electrostatic breakdown of the STO, which includes an oxide barrier such as MgO.


\subsection*{Method of evaluating the Lyapunov exponent}

The Shimada-Nagashima method \cite{shimada79} was used to evaluate the Lyapunov exponent of STO \cite{yamaguchi19,taniguchi19PRB,akashi20,taniguchi20,taniguchi22}. 
Note that the Lyapunov exponent is the expansion rate of the difference between two solutions of an equation of motion with slightly different initial conditions. 
In this method, at a certain time $t_{0}$, we add a small perturbation to a solution of the Thiele equation $\mathbf{X}(t_{0})$ as $\mathbf{X}^{(1)}(t_{0})=\mathbf{X}(t_{0})+\epsilon R \mathbf{n}_{0}$, where a unit vector $\mathbf{n}_{0}$ in the $xy$ plane points in an arbitrary direction and the meaning of the superscript ``(1)'' will be clarified below. 
The dimensionless parameter $\epsilon$ corresponds to the magnitude of the perturbation. 

We solve the Thiele equations of $\mathbf{X}(t_{0})$ and $\mathbf{X}^{(1)}(t_{0})$ and obtain $\mathbf{X}(t_{0}+\Delta t)$ and $\mathbf{X}^{(1)}(t_{0}+\Delta t)$, where $\Delta t$ is the time increment of the Thiele equation. 
Then, from the difference between $\mathbf{X}(t_{0}+\Delta t)$ and $\mathbf{X}^{(1)}(t_{0}+\Delta t)$, we evaluate the temporal Lyapunov exponent at $t_{1}=t_{0}+\Delta t$ as $\lambda^{(1)}=(1/\Delta t)\ln [\epsilon^{(1)}/\epsilon]$, where $\epsilon^{(1)}= |\mathbf{X}^{(1)}(t_{1})-\mathbf{X}(t_{1})|/R$. 

Next, we introduce $\mathbf{X}^{(2)}(t_{0}+\Delta t)$, which can be obtained by moving $\mathbf{X}(t_{0}+\Delta t)$ in the direction of $\mathbf{X}^{(1)}(t_{0}+\Delta t)$ over a distance $\epsilon$. 
Solving the Thiele equation of $\mathbf{X}(t_{0}+\Delta t)$ and $\mathbf{X}^{(2)}(t_{0}+\Delta t)$ yields $\mathbf{X}(t_{0}+2 \Delta )$ and $\mathbf{X}^{(2)}(t_{0}+2\Delta t)$. 
Then, the temporal Lyapunov exponent at $t_{2}=t_{0}+2\Delta t$ is obtained as $\lambda^{(2)}=(1/\Delta t)\ln[\epsilon^{(2)}/\epsilon]$ with $\epsilon^{(2)}=|\mathbf{X}^{(2)}(t_{2})-\mathbf{X}(t_{2})|/R$. 

Now we repeat the above procedure: i.e., first, we introduce $\mathbf{X}^{(n+1)}(t_{0}+n \Delta t)$ by moving $\mathbf{X}(t_{0}+n \Delta t)$ in the direction of $\mathbf{X}^{(n)}(t_{0}+\Delta t)$ over a fixed distance $\epsilon$. 
Second, by solving the Thiele equation, we obtain $\mathbf{X}^{(n+1)}[t_{0}+(n+1)\Delta t]$ and $\mathbf{X}[t_{0}+(n+1)\Delta t]$. 
Third, the temporal Lyapunov exponent at $t_{n+1}=t_{0}+(n+1)\Delta t$ is obtained as 
\begin{equation}
  \lambda^{(n+1)}
  =
  \frac{1}{\Delta t}
  \ln
  \frac{\epsilon^{(n+1)}}{\epsilon}, 
\end{equation}
where 
\begin{equation}
  \epsilon^{(n+1)}
  =
  \frac{|\mathbf{X}^{(n+1)}(t_{n+1})-\mathbf{X}(t_{n+1})|}{R}. 
\end{equation}
The Lyapunov exponent is the average of the temporal Lyapunov exponents defined as 
\begin{equation}
  \lambda
  =
  \lim_{N \to \infty}
  \frac{1}{N \Delta t}
  \sum_{i=1}^{N}
  \ln
  \frac{\epsilon^{(i)}}{\epsilon}. 
\end{equation}
Strictly speaking, the Lyapunov exponent obtained by this procedure corresponds to the maximum (or largest) Lyapunov exponent, i.e., the largest expansion rate of the difference between two solutions with slightly different initial conditions. 
The method assumes that, even though an initial perturbation is given in an arbitrary direction ($\mathbf{n}_{0}$ mentioned above), the perturbed solution will naturally move in the most expandable direction. 
In addition, the Lyapunov exponent evaluated here is a conditional one, where, although the system is non-autonomous because time-dependent input signals are injected, only the expansion rate of $\mathbf{X}$ is evaluated because the input signal is constant during a pulse width longer than the time increment \cite{akashi20}. 


\subsection*{Method of evaluating short-term memory capacity}

The memory capacity is a figure of merit quantifying the computational capability of physical reservoir computing \cite{dambre12}. 
As can be seen below, roughly speaking, the memory capacity is the number of the previous input signal which can be reproduced by the output signal from physical system. 
Let us assume that the input signals are pulse-shaped random number $q_{k}$, where the suffix $k=1,2,\cdots$ corresponds to the order of the input signals. 
In the present work, a binary input $b_{k}=0,1$ is used to evaluate the short-term memory capacity in accordance with Refs. \cite{tsunegi18JJAP,yamaguchi20,tsunegi19}. 
Note that a uniformly distributed random input $r_{k}$  ($-1 \le r_{k} \le 1$) has also been used in some reports \cite{akashi20}. 
An example of the target data is 
\begin{equation}
  y_{k,D}^{\rm STM}
  =
  q_{k-D}, 
  \label{eq:target_data_STM}
\end{equation}
where an integer $D$, called a delay, characterizes the order of the past input data. 
There are other kinds of target data, which, in general, include nonlinear transformation of input data \cite{kubota21}. 
A memory capacity is defined for each kind of target data. 
The short-term memory capacity is for target data given by Eq. (\ref{eq:target_data_STM}). 
The following procedure for evaluating of the short-term memory capacity is applicable to general target data. 
Therefore, for generality, let us denote the target data as $y_{k,D}$, for a while. 


The procedure is as follows.  
In physical reservoir computing, the physical system consists of many bodies, corresponding to neurons. 
The output signal from the $i$th neuron with respect to the $k$th input signal $q_{k}$ is denoted as $u_{k,i}$. 
Note that, since the present system consists of a single STO, we apply the time-multiplexing method \cite{nakajima21} to introduce virtual neurons. 
The output signal from a vortex STO is proportional to the $y$ component of the vortex-core position through the magnetoresistance effect. 
In experiments \cite{torrejon17,tsunegi19,tsunegi21}, the output signal is decomposed into the amplitude $s$ and the phase through the Hilbert transformation. 
The amplitude is used as a dynamical response in Refs. \cite{torrejon17,tsunegi21}, while the phase is used in Ref. \cite{tsunegi19}. 
In this work, we use the amplitude as an output signal. 
The virtual neurons $u_{k,i}$ are then defined as $u_{k,i}= s[t_{0} + (k-1 + i/N_{\rm node})t_{\rm p}]$, where $t_{0}$ is the time at which the first input signal is injected. 
The number of virtual neurons is denoted as $N_{\rm node}$. 
We inject the input signal $q_{k}$ ($k=1,2,\cdots,N$) to the STO and evaluate $u_{k,i}$, where $N$ is the number of input signals. 
After that, we introduce a weight $w_{D,i}$ and evaluate its value to minimize the error, 
\begin{equation}
  \sum_{k=1}^{N}
  \left(
    \sum_{i=1}^{N_{\rm node}+1}
    w_{D,i}
    u_{k,i}
    -
    y_{k,D}
  \right)^{2}, 
\end{equation}
where $u_{k,N_{\rm node}+1}=1$ is a bias term. 
The process determining the weight is called learning. 
Note that a weight is introduced for each target data and delay. 

After the weight is determined, we inject a different series of input data $q_{n}^{\prime}$ ($n=1,2,\cdots,N^{\prime}$), where $N^{\prime}$ is the number of input data $q_{n}^{\prime}$ and is, in general, not necessarily the same as $N$. 
The response from the $i$th neuron with respect to the $n$th input signal $q_{n}^{\prime}$ is denoted as $u_{n,i}^{\prime}$. 
Next, we introduce the system output defined as 
\begin{equation}
  v_{n,D}^{\prime}
  =
  \sum_{i=1}^{N_{\rm node}+1}
  w_{D,i}
  u_{n,i}^{\prime}. 
\end{equation}
The target data defined from $q_{n}^{\prime}$ is denoted as $y_{n,D}^{\prime}$. 

If the learning is done well, the system output $v_{n,D}^{\prime}$ closely reproduces the target data $y_{n,D}^{\prime}$. 
The reproducibility is quantified by the correlation coefficient, 
\begin{equation}
  {\rm Cor}(D)
  =
  \frac{\sum_{n=1}^{N^{\prime}}\left( y_{n,D}^{\prime} - \langle y_{n,D}^{\prime} \rangle \right) \left( v_{n,D}^{\prime} - \langle v_{n,D}^{\prime} \rangle \right)}
    { \sqrt{ \sum_{n=1}^{N^{\prime}}\left( y_{n,D}^{\prime} - \langle y_{n,D}^{\prime} \rangle \right)^{2} \sum_{n=1}^{N^{\prime}} \left( v_{n,D}^{\prime} - \langle v_{n,D}^{\prime} \rangle \right)^{2}}}, 
\end{equation}
where $\langle \cdots \rangle$ represents the averaged value. 
When the system output $v_{n,D}^{\prime}$ completely reproduces the target data $y_{n,D}^{\prime}$, $|{\rm Cor}(D)|=1$. 
On the other hand, when $v_{n,D}^{\prime}$ is completely different from $y_{n,D}^{\prime}$, $|{\rm Cor}(D)|=0$. 
Therefore, the correlation coefficient ${\rm Cor}(D)$ quantifies the reproducibility of the input signal $D$ times before the present data. 

As mentioned, the short-term memory capacity is the memory capacity when the target data $y_{k,D}$ is given by Eq. (\ref{eq:target_data_STM}). 
The short-term memory capacity is defined as 
\begin{equation}
  C_{\rm STM}
  =
  \sum_{D=1}^{D_{\max}}
  \left[
    {\rm Cor}(D)
  \right]^{2}. 
  \label{eq:STM_capacity_definition}
\end{equation}
The maximum delay $D_{\rm max}$ is $20$; see also Supplementary Information for the dependence of ${\rm Cor}(D)$ on $D$.
Note that, in this work, the short-term memory capacity is defined from the correlation coefficient ${\rm Cor}(D)$ from $D=1$, as can be seen in Eq. (\ref{eq:STM_capacity_definition}), whereas, in other work \cite{akashi20}, the correlation coefficient for $D=0$ is included in the definition.

\section*{Acknowledgements}

Y.I. and T.T. are grateful to Shingo Tamaru for his valuable discussion on the energy efficiency of pulse-signal injection. 
This paper is based on results obtained in a project (Innovative AI Chips and Next-Generation Computing Technology Development/(2) 
Development of next-generation computing technologies/Exploration of Neuromorphic Dynamics towards Future Symbiotic Society) commissioned by NEDO. 
The work is also supported by JSPS KAKENHI Grant Number 20H05655. 


\section*{Author contributions statement}

Y.I. and T.T. designed the project with help from K.N. and S.T. 
K.N. supported the statistical analysis of Y.I. and T.T. 
Y.I. and T.T. developed the codes, performed the simulations, wrote the manuscript and prepared the figures. 
All authors contributed to discussing the results. 


\section*{Competing interests}

The authors declare no competing interests. 


\section*{Data availability}

The datasets used and/or analysed during the current study available from the corresponding author on reasonable request.


\section*{Additional information}

\textbf{Supplementary information} is available for this paper. 
\\
\textbf{Correspondence} and requests for materials should be addressed to T.T.





\end{document}